\documentstyle[12pt, bezier]{article}
\newcommand {\bz}{\mbox{\boldmath $\zeta $}}
\begin{document}
\thispagestyle{empty}
\begin{flushleft}
{\Large \bf INTERACTION OF CHARGED 3D SOLITON WITH COULOMB CENTER}
\end{flushleft} 
\hskip 2 cm
\vbox{ 
\hbox{\bf Yu.\ P.\ RYBAKOV} 
\hbox{\it Department of Theoretical Physics} 
\hbox{\it Russian Peoples' Friendship University} 
\hbox{\it 6, Miklukho-Maklay str., 117198 Moscow, Russia} 
\hbox{\it e-mail: rybakov@udn.msk.su}
\vskip 3mm 
\hbox{\bf B.\ SAHA} 
\hbox{\it Laboratory of Theoretical Physics} 
\hbox{\it Joint Institute for Nuclear Research, Dubna} 
\hbox{\it 141980 Dubna, Moscow region, Russia} 
\hbox{\it e-mail:  saha@thsun1.jinr.dubna.su}} 
\vskip 1.5cm
\noindent
The Einstein - de Broglie particle-soliton concept is applied to simulate 
stationary states of an electron in a hydrogen atom. According to this 
concept, the electron is described by the localized regular solutions 
to some nonlinear equations. In the framework of Synge model  for  
interacting  scalar  and electromagnetic fields a system of integral  
equations has been obtained, which describes the interaction between   
charged 3D soliton and Coulomb center. The  asymptotic  expressions  for 
physical fields, describing soliton moving around the fixed Coulomb 
center, have been obtained with the help of integral equations. 
It is shown that the electron-soliton center travels along some stationary 
orbit around the Coulomb center. The electromagnetic radiation is absent 
as the Poynting vector has non-wave asymptote $O(r^{-3})$ after averaging 
over angles, i.e. the existence of spherical surface corresponding to 
null Poynting vector stream, has been proved. Vector lines for Poynting 
vector are  constructed  in asymptotical area. 


\newpage 
\section{Introduction}

From the history of quantum mechanics it is known that as early as 1927 
 in the framework of his "theory of double solution" Louis de Broglie made 
an attempt to represent the electron as a source of waves 
obeying the Schr$\ddot o$dinger equation [1]. Later he modified his 
model showing that the electron should be described by regular 
solutions to some nonlinear equation coinciding with the Schr$\ddot 
o$dinger one in the linear approximation. This scheme became famous as a
causal nonlinear interpretation of quantum mechanics [2]. Developing this 
concept, de Broglie remarked that it had much in common with 
Einstein's ideas about unified field theory according to which particles 
were to be considered as clots of some material fields obeying the 
nonlinear field equations [3]. In recent years, these types of field 
configurations, known as soliton or particle-like solutions, came into 
active use to model extended elementary particles [4].

In this paper the Einstein-de Broglie soliton concept is employed to 
model stationary states of the electron in a hydrogen atom.

\section{Bohm problem about nonlinear resonance and its possible 
solution}
\setcounter{equation}{0}

As a starting point, we will consider an interesting problem posed by 
D.\ Bohm. Long ago, in his book [5] D.\ Bohm discussed the possible 
connection between the wave-particle dualism in quantum mechanics and 
the hypothetical nonlinear origin of fundamental equations in a 
future theory of elementary particles. To illustrate the line of Bohm's 
argument we will consider a simple scalar model in the Minkowski 
space-time given by the Lagrangian density \begin{equation} {\cal L}= 
{\partial}_i \phi^{*}\, {\partial}_j \phi\, \eta^{ij} - (mc/\hbar)^2 
\phi^{*}\phi + F(\phi^{*}\phi).  \end{equation} Here $i,j= 0,1,2,3; \quad 
\eta^{ij}= \mbox{diag}(1,-1,-1,-1)$, the nonlinear function $F(s)$ behaves 
at $s \to 0 $ as $s^n, \quad n>1$, and is assumed as such that the 
corresponding field equations allow the existence of particle-like 
(soliton) solutions, i.e. regular configurations localized in space and 
endowed with finite energy. In particular, it can be shown that if one 
chooses $F(s)=k s^{3/2}, \quad k>0$, the model (2.1), known as the Synge 
one [6], admits the following stationary solutions:  \begin{equation} 
\phi_0= u(r) \exp{(-i\omega_0 t)}, \quad r= |{\bf r}|.  \end{equation} 
Here, the real radial function $u(r)$ is regular everywhere and 
exponentially decreases as $r \to \infty$, that provides finiteness of 
energy of the configuration \begin{equation} E= \int d^3 x\, 
T^{00}(\phi_0), \end{equation} where $T^{ij}$ is the corresponding 
energy-momentum tensor.

Moreover the model mentioned is intriguing due to the fact that  
nodeless solitons turn out to be stable by Lyapunov provided that 
their charge is fixed [7]. So there exist perturbed solitons slightly 
differing from the stationary solitons (2.2):
\begin{equation}
\phi = \phi_0 + \xi(t, {\bf r}).
\end{equation}
Note that the perturbation $\xi$ in (2.4) is small as compared with 
$\phi_0$ only in the area of localization of the soliton, where $\phi_0$ 
significantly differs from zero. None the less far from the soliton 
center, where $\phi_0$ is negligibly small, one can put $\phi \approx 
\xi$, i.e.  the {\it tail} of the soliton is completely defined by the
perturbation $\xi$.

D.\ Bohm put the following question: Does there exist any nonlinear 
model for which the spatial asymptote (as $r \to \infty$) of a perturbed 
soliton-like solution represents oscillations with 
characteristic frequency $\omega=E/\hbar$? In other words, for the 
model in question the principal Fourier-amplitude in the expansion of 
the field $\phi \approx \xi$ as $r \to \infty$ should correspond to the
frequency $\omega$ connected with the soliton energy (2.3) by the 
Planck-de Broglie formula  \begin{equation} E=\hbar \omega.  
\end{equation}

Note that for the model (2.1) at spatial infinity, where $\phi \to 0$, 
the field equation reduces to the linear Klein-Gordon one
\begin{equation}
[\Box -(mc/\hbar)^2]\phi = 0,
\end{equation} 
and therefore the relation (2.5) holds only for solitons with 
unique energy $E=mc^2$ defined by the mass $m$ fixed in (2.1). Thus the 
universality of the relation (2.5) breaks down in the model (2.1), 
so forcing its modification. In the light of the above universality,  
the frequency $\omega$ in (2.5) being defined by the mass of 
the system, it seems natural that in the new, modified model 
one should use the gravitational field, spatial asymptote of which is 
also defined by the mass of the considered localized system. Thus, to 
solve the Bohm problem the possibility to invoke the gravitational field 
comes into reality [8].

So we will describe the new model with the Lagrangian density ${\cal L}= 
{\cal L}_m +{\cal L}_g$, where $${\cal L}_g = c^4 R/16\pi G$$ corresponds 
to the Einstein's theory of gravity, and ${\cal L}_m$ is chosen as 
\begin{equation} {\cal L}_m= {\partial}_i \phi^{*}\, {\partial}_j \phi\, 
g^{ij} - I(g_{ij})\phi^{*}\phi + F(\phi^{*}\phi).  \end{equation} The 
crucial point of this scheme is to build up the invariant $I(g_{ij})$ 
depending on the metric tensor $g_{ij}$ of the Riemannian space-time and 
its derivatives.  This invariant should possess such properties that in 
the vicinity of the soliton with mass $m$, the relation \begin{equation} 
\lim\limits_{r \to \infty} I(g_{ij})= (mc/\hbar)^2  \end{equation} should 
hold.  It is easy to see that on the basis of (2.8) one can  
asymptotically deduce the equation (2.6) from the Lagrangian (2.7).

We argue that the invariant $I$ can be built from the curvature 
tensor $R_{ijkl}$  and its covariant derivatives $R_{ijkl;n}$:  
\begin{equation}
I=(I_{1}^{4}/I_{2}^{3})c^6\hbar^{-2}G^{-2},
\end{equation}
where $G$ is the Newton's gravitational constant and the invariants $I_1$ 
and $I_2$ take the form $$ I_1=R_{ijkl}R^{ijkl}/48, \quad 
I_2=-R_{ijkl;n}R^{ijkl;n}/432.$$ Estimating $R^{ijkl}$ at large distance 
$r$ with the help of the Schwarzschild metric, one can find $$ I_1=G^2 
m^2/(c^4 r^6), \quad I_2=G^2 m^2/(c^4 r^8).$$ So from (2.9) there 
immediately follows (2.8).  Thus, within the modified model (2.7) for all 
massive particles the Planck-de Broglie relation (2.5) is automatically 
fulfilled. It means that in the framework of the scheme mentioned the 
principle of wave-particle dualism is valid, according to which the 
relation (2.5) is realized as a condition of the nonlinear resonance.

To verify the fact that solitons can really possess wave 
properties, the {\it gedanken} diffraction
experiment with individual electron-solitons similar to the numerical one 
of Biberman et al. [9], was realized. Solitons with some velocity 
were dropped into a rectilinear slit, cut in the impermeable screen, and 
the transverse momentum was calculated which they gained while passing the 
slit the width of which significantly exceeded the size of the soliton. As 
 a result, the picture of distribution of the centers of scattered 
solitons was restored on the registration screen, by considering their 
initial distribution to be uniform over the transverse coordinate. It was 
clarified that though the center of each soliton fell into a definite 
place of the registration screen (depending on the point of crossing of 
the slit and the initial soliton profile), the statistical picture in many 
ways was similar to the well-known diffraction distribution in optics, 
i.e.  Fresnel's picture at short distances from the slit and Fraunhofer's 
picture at large distances [10,11].

Fulfillment of the quantum mechanics correspondence principle for the 
Einstein-de Broglie's soliton model was discussed in the works [12-15]. In 
these papers it was shown that in the framework of the soliton model all 
quantum postulates were regained at the limit of point particles so that 
from the physical fields one can build the amplitude of probability and 
the average can be calculated as a scalar product in the Hilbert 
space by introducing the corresponding quantum operators. In this paper, 
we will show that in the framework of the Einstein-de Broglie soliton 
model a hydrogen atom can be simulated.

\section{Fundamental equations and structure of solutions}
\setcounter{equation}{0}

\begin{minipage}{50mm}
Let us consider the hydrogen atom with the electron replaced by a 
localized object "soliton" that is moving round the nucleus. So that the 
soliton-like solution does exist one has to consider a nonlinear model.
\end{minipage}
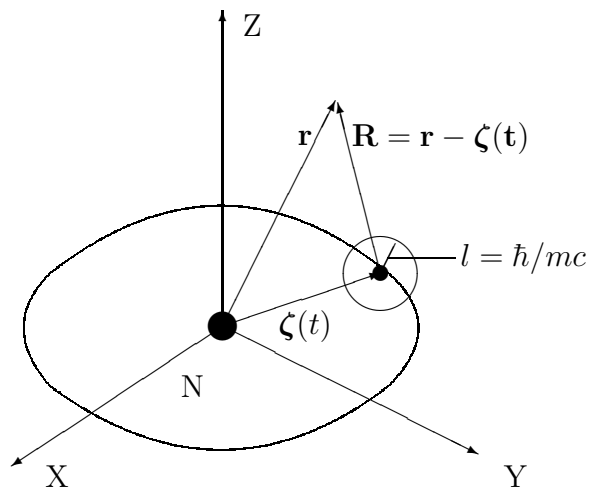
\begin{minipage}{100mm}
\unitlength=1.00mm
\special{em:linewidth 0.4pt}
\linethickness{0.4pt}
\begin{picture}(98.00,95.00)
\put(58.00,53.00){\circle*{4.00}}
\bezier{232}(35.00,60.00)(58.00,78.00)(80.00,60.00)
\bezier{84}(35.00,60.00)(28.00,53.00)(35.00,45.00)
\bezier{224}(35.00,45.00)(58.00,28.00)(80.00,45.00)
\bezier{88}(80.00,45.00)(88.00,53.00)(80.00,60.00)
\put(58.00,53.00){\vector(0,1){42.00}}
\put(58.00,53.00){\vector(2,-1){34.00}}
\put(58.00,53.00){\vector(-3,-2){28.00}}
\put(58.00,53.00){\vector(3,1){21.00}}
\put(79.00,60.00){\circle*{2.00}}
\put(79.00,60.00){\circle{10.00}}
\put(54.00,45.00){\makebox(0,0)[cc]{N}}
\put(36.00,33.00){\makebox(0,0)[cc]{X}}
\put(97.00,33.00){\makebox(0,0)[cc]{Y}}
\put(62.00,93.00){\makebox(0,0)[cc]{Z}}
\put(69.00,78.00){\makebox(0,0)[cc]{$\bf r$}}
\put(69.00,53.00){\makebox(0,0)[cc]{$\bz(t)$}}
\put(87.00,78.00){\makebox(0,0)[cc]{$\bf R=\bf r-\bz(t)$}}
\put(98.00,62.00){\makebox(0,0)[cc]{$l=\hbar/mc$}}
\put(79.00,60.00){\line(1,2){2.00}}
\put(80.00,62.00){\line(1,0){9.00}}
\put(79.00,60.00){\vector(-1,4){5.67}}
\put(58.00,53.00){\vector(1,2){15.00}}
\end{picture}
\end{minipage}

As physical fields we choose the complex scalar field $\phi$ 
interacting with the electromagnetic one $F_{ik}=\partial_i A_k - 
\partial_k A_i$. The nucleus field is assumed to be  the
Coulomb one: $A_{i}^{ext}=\delta_{i}^{0}Ze/r$. The 
Lagrangian density is taken in the following form
\begin{equation}
{\cal L}=-\frac{1}{16\pi}(F_{ik})^2 + |[\partial_k - i\epsilon(A_k + 
A_{k}^{ext})]\phi|^2 - (mc/\hbar)^2\phi^*\phi + F(\phi^* \phi),
\end{equation}
where $\epsilon=e/(\hbar c)$ is the coupling constant, $F(\phi^* \phi)$ 
is some nonlinear function, decreasing faster as $\phi \to 0$ than 
$|\phi|^2$ and is chosen so that the field equation at 
$A_{i}^{ext}=0$ allows the existence of stable stationary soliton-like 
solutions of the type (2.2), describing configurations with mass $m$ 
and charge $e$.

Note that for simplicity we do not write down the terms corresponding to 
the gravitational field that will be taken into account implicitly with 
the help of the nonlinear resonance condition (2.5).

Let us consider the nonrelativistic approximation assuming that 
\begin{equation}
\phi = \psi\,  \exp{(-imc^2t/\hbar)},
\end{equation} 
neglecting in the equations of motion higher derivatives of $\psi$ 
with respect to time and retaining only linear terms in $A_i$. As a 
result, taking (3.2) into account we get the following system of 
equations:  
\begin{equation} 
i\hbar\, \partial_t \psi + (\hbar^2/2m)\triangle \psi +(Ze^2/r)\psi\,
=\, -(\hbar^2/2m) {\hat f}({\bf A}, A_0, \psi^*\psi)\psi, 
\end{equation} 
\begin{equation} 
\Box A_0=(8\pi me/\hbar^2)|\psi|^2 \equiv 
-4\pi \rho, 
\end{equation} 
\begin{equation} \Box {\bf A}=4\pi [2\epsilon^2 
{\bf A} |\psi|^2 - i\epsilon(\psi^* \nabla \psi - \psi \nabla \psi^*)] 
\equiv -(4\pi /c)\, {\bf j},  
\end{equation} 
where
$$ {\hat f}({\bf A}, A_0, \psi^*\psi)\psi\,\equiv\,
2i\epsilon({\bf A} \nabla)\psi + 2(\epsilon 
mc/\hbar) A_0 \psi +i\epsilon \psi\, \mbox{div} {\bf A} + 
F^{\prime}(\psi^* \psi)\psi. $$ 
Moreover in the equations 
(3.3 - 3.5) it is supposed that the 4-potential $A_i$ of the proper 
electromagnetic field of the soliton obeys to the Lorentz condition 
$$\partial_t A_0 + \mbox{c div} {\bf A} =0,$$ which is consistent with  
equations (3.3) - (3.5) owing to the conservation of electric charge.

We will seek for the solutions to equations (3.3) - (3.5) describing 
the stationary state of an atom when the electron-soliton center is 
assumed to be moving along a circular orbit of the radius $a_0$ with some 
angular velocity $\Omega$. In this problem there arise two characteristic 
lengths: the size of the soliton $l=\hbar/(mc)$ and the Bohr's radius 
$a=\hbar^2/(mZe^2).$ It is obvious that $a_0 \sim a  \gg l$.

Let us first consider the area near the soliton center where $r-a_0 \sim 
l$. Suppose the soliton center trajectory to be ${\bf r} = {\bz}(t)$. 
Putting into (3.3) the configuration
$$ \psi = u({\bf r} - {\bz}(t))\, \exp{(i{\cal S}/\hbar)},$$
neglecting the contribution of the proper electromagnetic field and 
separating the real and imaginary parts, we get   
\begin{equation}
\partial_t {\cal S} - \frac{Ze^2}{r} + \frac{1}{2m}(\nabla {\cal S})^2 - 
\frac{\hbar^2}{2m}(\hat f +\frac{\triangle u}{u})=0, 
\end{equation} 
\begin{equation}
\triangle {\cal S} +2 (\nabla {\cal S} - m \dot {{\bz}}) \cdot 
\nabla u /u =0.  \end{equation} Assuming $S$ to be a slowly varying 
function of a point in the vicinity of the soliton center, from (3.7) we 
deduce \begin{equation} {\cal S} \approx m \dot {{\bz}} \cdot ({\bf 
r} - {\bz}) + C_0t +\chi (t), \quad C_0=\mbox{const}.  
\end{equation}
Taking into account the classical equations of motion of a charged 
particle in the Coulomb field $$m \ddot {{\bz}} = -Ze^2 {\bz 
}/\zeta^3 $$ and using the expansion $$ \frac{1}{r} \approx 
\frac{1}{\zeta} - \frac{{\bz} \cdot ({\bf r} - {\bz})}{\zeta^3}, 
$$ from (3.6) and (3.8) we derive $$\partial_t \chi = \frac{m}{2} {\dot 
{{\bz}}}^2 + \frac{Ze^2}{\zeta} \equiv {\cal L}(t), $$ where 
${\cal L}(t)$ is the Lagrangian of a particle in the Coulomb field. Thus, 
the function $\chi$ is the classical action on the trajectory:  
\begin{equation} \chi(t)= \int\limits_{0}^{t} {\cal L}(t)dt, 
\end{equation} and the function $u$ is the soliton-like solution to the 
quasi-stationary problem \begin{equation} \hbar^2 (\hat f + \triangle u 
/u)= 2mC_0.  \end{equation} In this case according to (3.4) and (3.5) $$ 
\rho = -(2me/\hbar^2) u^2, \quad {\bf j} = - 2\epsilon c u^2 (\epsilon 
{\bf A} + m \dot {{\bz}}/\hbar), $$ which makes it possible, using 
the common solutions to equations (3.4), (3.5) and (3.10), to calculate 
the potentials $A_i$ of the electromagnetic field in the vicinity of the 
soliton center:  $$ A_0 = A_0({\bf r} - {{\bz}}(t)), \quad c {\bf A} 
= {\dot {{\bz}}}(t)A_0({\bf r} - {{\bz}}(t)),$$ where the 
terms ${\dot {{\bz}}}^2/c^2$ are neglected.

Let us now study the asymptotic behavior of soliton 
at large distance, i.e. we will study it's {\it "tail"}. We will use  
successive approximation method. Thus, we  need to  rewrite  the  
differential equations (3.3), 
(3.4) and (3.5) in integral form. Note that we are not  solving  Cauchy 
evolution problem, choosing definite initial condition.  
Were it a question of  Cauchy  problem,  we  would  bound  to  use 
retarded Green's function. In this case it has been assumed  that the 
object under consideration (atom) already  exists  infinitely  long.  
Thus we consider the problem to study corresponding steady state, which 
eliminates  the possibility  of using  retarded electromagnetic Green's 
function.  The problem mentioned, according  to  our view,  can  be 
satisfied by half-sum of retarded and advanced  solutions.  Above 
mentioned selection can  be  justified  by the  assumption  that during 
the evolution process the  radiation  of  the independent i.e. 
half-difference of retarded and advanced electromagnetic fields took place.

To find out the field $\psi$ far from the soliton center, we  
rewrite equation (3.3) in the integral form
\begin{eqnarray}
\psi (t, {\bf r}) &=& C_n \psi_n({\bf r})\, \exp{(-i\omega_nt)} \, + 
\nonumber \\
&+& \frac{1}{2\pi}\int d\omega \int dt^{\prime} \int d^3x^{\prime}\, 
\exp{[-i\omega(t-t^{\prime})]}\,G({\bf r}, {\bf r}^{\prime};\omega+i0) 
\hat f \psi (t^{\prime}, {\bf r}^{\prime}), 
\end{eqnarray} 
where $\psi_n({\bf r})$ is 
the eigenfunction of the Hamiltonian of a hydrogen atom for a stationary 
state of number $n$ with energy $E_n=\hbar \omega_n, \quad C_n 
=\mbox{const}$ and $G({\bf r}, {\bf r}^{\prime};\omega)$ is the 
Hamiltonian's resolvent having the form [16] \begin{equation} G({\bf r}, 
{\bf r}^{\prime};\omega) = \frac{\Gamma (1-i\nu)}{4\pi R} 
\left|\begin{array}{ccc}W_{i\nu,1/2}(-ikr_+)&M_{i\nu,1/2}(-ikr_-)\\
{\dot W}_{i\nu,1/2}(-ikr_+)&{\dot M}_{i\nu,1/2}(-ikr_-) \end{array}\right|.
\end{equation}    
Here,the following notation is used:
$$k=(2m\omega/\hbar)^{1/2}, \quad Im\, k>0, \quad \nu=(ka)^{-1}, $$
$$r_{\pm} =r + r^{\prime} \pm |{\bf r} - {{\bf r}}^{\prime}|$$
and the Whittaker functions $ W_{i\nu,1/2}, M_{i\nu,1/2}$ and their 
derivatives ${\dot W}_{i\nu,1/2}$, ${\dot M}_{i\nu,1/2}$ are introduced.
To find the field $\psi$ at large distances from the electron-soliton 
center, i.e. at $|r - a_0| \gg l$, it is sufficient to put in (3.11)
\begin{equation}
\hat f\, \psi (t, {\bf r})= g\, \exp{(-i\omega_nt)}\,\delta({\bf r} - {\bz 
}(t)), \quad g=const,
\end{equation}
where the relation (2.5) is taken into account. As a result, we get
\begin{eqnarray}
\psi (t, {\bf r}) &=& C_n \psi_n({\bf r})\, \exp{(-i\omega_nt)} \, +
\nonumber \\
&+& \frac{1}{2\pi}\int d\omega \int dt^{\prime}\,  
\exp{[-i\omega\, t+it^{\prime}(\omega- \omega_n)]}\,G({\bf r}, {\bf 
r}^{\prime};\omega+i0). 
\end{eqnarray}
It is easy to verify that the field (3.14) decreases exponentially at 
large distances.
With the help of (3.14) and equations (3.4) and (3.5), one can evaluate 
the electromagnetic field outside the soliton. 
In (3.14) $C_n$ is unknown constant and $\psi_n({\bf r})$ is the wave 
function of electron in stationary state. In each step of iteration one 
obtains (3.14) where the stationary "tail" of soliton is marked out and 
its center moves along some effective orbit. The orbital parameter and 
constant $C_n$ may be defined in arbitrary approximation of 
minimization $|| \psi_{(k)}-\psi_{(k+1)}||$ [17]. Still now the 
constants $C_n$ and $g$ are not found explicitly and we hope to obtain 
them in our forthcoming papers.

Let us now solve the equations (3.4) and 
(3.5).  Considering that the nonlinear source is rather weak one, we will 
replace the right hand sides of the equations (3.4) and (3.5) by 
$\delta$- functions. Let us also notice that 
$${\bf E}_{-}\,=\,({\bf 
E}_{-}+{\bf E}_{+})/2\,+\, ({\bf E}_{-}-{\bf E}_{+})/2,$$ 
where ${\bf 
E}_-= {\bf E}^{ret}, \quad {\bf E}_+= {\bf E}^{adv}.$ It is well known 
that the half-difference of retarded and advanced fields radiates.  So it 
will be sufficient to consider, as was discussed earlier, the half-sum  
of retarded and advanced solutions, describing the  stationary state.  It 
means,  we will seek the strength of the electromagnetic field as the 
half-sum of those for retarded and advanced fields. It means  that for 
large times $|\omega_n|t \gg 1$ the 4-potential $A_k$ will contain only 
stationary part $A_k=(A_{k}^{ret}+ A_{k}^{adv})/2$. 

Let us find the expression for ${\bf E}_{-}$. Radius of the soliton 
$l$ is rather small in comparison to Bohr radius $a$, i.e. $a\gg l$.  
So the source can be considered as proper one. Let the point-like 
charge $e$ move  along  the  given  trajectory $\bf r=\bz(t)$  with  
velocity $\bf v(t)=\dot{\bz}(t).$  Then, to describe the electromagnetic 
field, generated by the charge, one can write charge density and 
current density as
\begin{equation}
\rho(t, {\bf r})= e\,\delta[{\bf r} - \bz(t)], \quad
{\bf j}(t, {\bf r})= e  {\bf v}(t)\,\delta[{\bf r} - \bz(t)].
\end{equation}
Then the equations (3.4) and (3.5) take form: 
\begin{eqnarray}
\Box A_0&=& -4\pi e\,\delta[{\bf r} - \bz(t)],  \nonumber \\
\Box {\bf A}&=& -\frac{4\pi e {\bf v}}{c}\,\delta[{\bf r} - 
\bz(t)].\nonumber
\end{eqnarray}
that lead to the well-known Lienard- Wiechert 
potentials. As we know, the retarded time  writes
$$
t_{-}=t- R(t_{-}, {\bf r})/c, 
$$                 
which leads to 
\begin{eqnarray}
\frac{\partial t_{-}}{\partial t}=\frac{1}{1 - ({\bf n}_{-}\cdot{\bf 
v})/c}, \quad  
\frac{\partial {\bf R}(t_{-}, {\bf r})}{\partial t_{-}}=
- ({\bf n}_{-}\cdot{\bf v}), \quad
\nabla t_{-}=
-\frac{{\bf n}_{-}}{c -({\bf n}_{-}\cdot{\bf v})}. \nonumber  
\end{eqnarray}
where ${\bf n}\,=\,{\bf R}(t,\,{\bf r})/R(t,\,{\bf r}).$
Later, using the following expressions for $A_{0-}(t, {\bf r})$
and ${\bf A_{-}}(t, {\bf r})$  [18,\,19]
$$
A_{0-}(t, {\bf r})=\frac{ec}{[cR - ({\bf v}\cdot {\bf R})]_{t_{-}}}, 
\quad 
{\bf A}_{-}(t, {\bf r})\Biggl[
\frac{e{\bf v}}{cR - ({\bf v}\cdot {\bf R})}\Biggr]_{t_{-}},$$
one finds the expression for retarded field
\begin{equation}
{\bf E}_{-}=\,e\Biggl[\frac{(c^2-v^2)(c{\bf n}_{-}-{\bf v})}
{R^2 [c -({\bf n}_{-}\cdot{\bf v})]^3}\,+\,
\frac{[{\bf n}_{-}, [(c{\bf n}_{-}-{\bf v}) , \dot {\bf v}]]}
{R [c -({\bf n}_{-}\cdot{\bf v})]^3}\Biggr]_{t_{-}},
\end{equation}
and
\begin{equation}
{\bf B}_{-}\,=\,\bigl[{\bf n}_{-}, {\bf E}_{-} \bigr]_{t_{-}}.
\end{equation}
Analogically writing the advanced time as
$$t_{+}\,=\,t+ R(t_{+}, {\bf r})/c,  $$
for the advanced field we find
\begin{equation}
{\bf E}_{+}=\,e\Biggl[\frac{(c^2-v^2)(c{\bf n}_{-}+{\bf v})}
{R^2 [c +({\bf n}_{-}\cdot{\bf v})]^3}\,+\,
\frac{[{\bf n}_{-}, [(c{\bf n}_{-}+{\bf v}), \dot {\bf v}]]}
{R [c +({\bf n}_{-}\cdot{\bf v})]^3}\Biggr]_{t_{-}},
\end{equation}
and
\begin{equation}
{\bf B}_{-}\,=\,-\bigl[{\bf n}_{+}, {\bf E}_{+} \bigr]_{t_{+}}.
\end{equation}
To calculate the power, lost by the charge due 
to radiation, one  has  to  compose  Poynting  vector ${\bf S}$ and 
retain the terms of the order 
$1/R^2$  as  the  integration  will take place along infinitely distant 
surface. As we know, Poynting vector is expressed by the relation 
\begin{equation}
{\bf S}\,=\,\frac{c}{4 \pi}[{\bf E},{\bf B}].
\end{equation}
In this case the field strengths are 
\begin{equation} 
{\bf E} = \frac{1}{2}({\bf E}_{+} + {\bf E}_{-}), \quad {\bf B} = 
\frac{1}{2}([{\bf n}_{-}, {\bf E}_{-}] - [{\bf n}_{+}, {\bf E}_{+}]), 
\end{equation} 
where ${\bf E}_-= {\bf E}^{ret}, \quad {\bf E}_+= 
{\bf E}^{adv}, \quad {\bf n}_{\pm}= {\bf R}_{\pm}/R_{\pm}, \quad {\bf 
R}_{\pm}= {\bf r} - {\bz}(t_{\pm})$ and $t_{\pm}$ are the roots of 
the equations $t_{\pm}= t \pm R_{\pm}/c$.

\noindent
Using some manipulations from vector analysis we rewrite 
Poynting vector as \begin{eqnarray} {\bf S}=\frac{c}{16 \pi}\bigl\{{\bf 
n}_{-}{\bf E}_{-}^{2} - {\bf n}_{+}{\bf E}_{+}^{2} +({\bf n}_{-}-{\bf 
n}_{+})({\bf E}_{-}\cdot {\bf E}_{+})-  \nonumber\\ -{\bf E}_{-}(({\bf 
E}_{-}+{\bf E}_{+})\cdot {\bf n}_{-}) + {\bf E}_{+}(({\bf E}_{-}+{\bf 
E}_{+})\cdot {\bf n}_{+})\bigr\}.  \end{eqnarray} Let us now rewrite all 
these in spherical  system  of coordinates. In this case \begin{eqnarray} 
{\bf R}({\bf r}, t_{\pm})&=&\{r-a_0\,\mbox{sin}\theta\, 
\mbox{cos}\alpha_{\pm},\, -a_0\,\mbox{cos}\theta\,
\mbox{cos}\alpha_{\pm},\, a_0\, \mbox{sin}\alpha_{\pm}\}, \nonumber \\ 
{\bf 
v}(t_{\pm})&=&\{a_0\,\Omega\,\mbox{sin}\theta\,\mbox{sin}\alpha_{\pm}, \, 
a_0\,\Omega\,\mbox{cos}\theta\,\mbox{sin}\alpha_{\pm},\, a_0\, \Omega\, 
\mbox{cos}\alpha_{\pm}\}, \nonumber \\ 
\dot{\bf v}(t_{\pm})&=&\{-a_0\,\Omega^2\,\mbox{sin}\theta\, 
\mbox{cos}\alpha_{\pm},\, -a_0\,\Omega^2\,\mbox{cos}\theta\, 
\mbox{cos}\alpha_{\pm}, \,a_0\,\Omega^2\,\mbox{sin}\alpha_{\pm}\}, 
\nonumber \\ R({\bf 
r},t_{\pm})&=&\sqrt{r^2\,+\,a_{0}^{2}\,-\,2a_0r\,\mbox{sin}\theta\,\mbox{cos} \alpha_{\pm}},\nonumber
\end{eqnarray} where $\alpha_{\pm}\,=\,\alpha-\Omega\, t_{\pm}$.  
Retaining the terms of the order $a_0/r$ for ${\bf n}_{-}$  and ${\bf 
n}_{+}$ one can obtain \begin{eqnarray} {\bf n}_{\pm}=\{1, -(a_0/r)\, 
\mbox{cos}\theta\,\mbox{cos}(\alpha - \Omega t_{\pm}), (a_0/r)\, 
\mbox{sin}(\alpha - \Omega t_{\pm})\}. \nonumber \end{eqnarray} Using the 
first nonvanishing approximation of the order $v/c$, from (3.16) and 
(3.18) we will get \begin{eqnarray} {\bf E}_{-}\approx e\Bigl[\frac{{\bf 
n}_{-}}{R^2}+ \frac {{\bf n}_{-}({\bf n}_{-}\cdot\dot{\bf v}) - \dot{\bf 
v}}{c^2 R}\Bigr]_{t_{-}}, \\ {\bf E}_{+}\approx e\Bigl[\frac{{\bf 
n}_{+}}{R^2}+ \frac {{\bf n}_{+}({\bf n}_{+}\cdot\dot{\bf v}) - \dot{\bf 
v}}{c^2 R}\Bigr]_{t_{+}}. \end{eqnarray} Taking into account that 
\begin{eqnarray}
1/R\approx 1/r - (a_0/r^2)\mbox{sin}\theta\,\mbox{cos}\alpha_{-}, \quad
1/R^2 \approx 1/r^2, \nonumber \\
({\bf n}_{-}\cdot \dot {\bf v})\approx -a_0 \Omega^2\,
\mbox{sin}\theta\,\mbox{cos}\alpha_{-} +(a_{0}^{2}\Omega^2 /r)[1-
\mbox{sin}^2\theta\, \mbox{cos}^2\alpha_{-}], \nonumber \
\end{eqnarray}
where $\alpha_{-}=\alpha-\Omega t_{-}$, for ${\bf E}_{-}$ one gets
\begin{eqnarray}
{\bf E}_{-}\approx e\bigl\{\frac{1}{r^2}[1+ (a_{0}^{2}\Omega^2 /c^2)(1-
\mbox{sin}^2\theta\,\mbox{cos}^2\alpha_{-})],
(a_0\Omega^2 /c^2 r)\,\mbox{cos}\theta\, \mbox{cos}\alpha_{-},\,
-(a_0\Omega^2 /c^2 r)\,\mbox{sin}\alpha_{-}\bigr\}.  \nonumber
\end{eqnarray}
Analogically one finds 
\begin{eqnarray}
{\bf E}_{+}\approx e\bigl\{\frac{1}{r^2}[1+ (a_{0}^{2}\Omega^2 /c^2)(1-
\mbox{sin}^2\theta\, \mbox{cos}^2\alpha_{+})],
(a_0\Omega^2 /c^2 r)\,\mbox{cos}\theta\, \mbox{cos}\alpha_{+},\,
-(a_0\Omega^2 /c^2 r)\,\mbox{sin}\alpha_{+}\bigr\}. \nonumber
\end{eqnarray}
Putting the above expressions for ${\bf n}_{\pm}, {\bf E}_{\pm}$ into 
(3.22) one can find the expressions for Poynting vector.
In doing so we will take into account that the normals
${\bf n}_{-}$ and ${\bf n}_{+}$ coincide as $r \to \infty$ with 
${\bf n} = {\bf r}/r$. Retaining the terms $(a_0/r)^3$ and also taking 
into account that $a_0\Omega =v \ll c$, for the circular motion in the 
spherical coordinates $r, \theta$ and $\alpha$ we have the following 
components of the Poynting vector ${\bf S}$:  \begin{eqnarray} 
S_r&=&\frac{e^2\,a_{0}^{2}\,\Omega^4}{16 \pi c^3\,r^2}\,
\mbox{sin}^2\theta\, \mbox{sin} 2(\alpha - \Omega 
t)\, \mbox{sin} (2\Omega r/c),  \nonumber \\
S_\theta &=& \frac{e^2\,a_0\,\Omega^2}{4 \pi c\,r^3}\, 
\mbox{cos}\theta\, \mbox{sin} (\alpha - \Omega 
t)\, \mbox{sin} (\Omega r/c),  \nonumber \\
S_\alpha &=& \frac{e^2\,a_0\,\Omega^2}{4 \pi c\,r^3}\,
\,\mbox{cos} (\alpha - \Omega 
t)\, \mbox{sin} (\Omega r/c).  
\end{eqnarray} 
It is obvious that for $r_k\,=\,ck\pi/\Omega$ with $k=0,1,2,\cdots$ all 
the components of Poynting  vector  turn  to zero i.e. ${\bf S}\,=\, 0.$  

All the calculations done above can be summed up as follows.
From (3.21) it follows that the projection of the Poynting vector ${\bf 
S}$ in the direction of the vector ${\bf N}=({\bf n}_+ + {\bf n}_-)/2 $, 
coinciding as $r \to \infty$ with ${\bf n} = {\bf r}/r$, takes the form 
\begin{equation} S_N=\frac{c}{16\pi \sqrt{2}}(E_{-}^{2} - E_{+}^{2})(1+ 
{\bf n}_+ \cdot {\bf n}_-)^{1/2}.  \end{equation} Since ${\bf n}_{\pm}= 
{\bf n} + O(r^{-1})$, after averaging expression (3.26) over the sphere, 
we find \begin{equation} <S_N> = \frac{c}{16\pi}\biggl(<E_{-}^{2}> - 
<E_{+}^{2}>\biggr) = O(r^{-3}).  \end{equation}

Thus according to (3.27) the electromagnetic radiation from the system 
is absent. In particular, for the circular motion in the spherical 
coordinates $r, \theta$ and $\alpha$ we have the following structure 
of the Poynting vector ${\bf S}$:
\begin{eqnarray}
S_r &=& \frac{\kappa}{r^2}\,\mbox{sin}^2\theta\, \mbox{sin} 2(\alpha - 
\Omega t)\, \mbox{sin} (2\Omega r/c), \\ \nonumber S_\theta &=& 
\mbox{sin}(\Omega r/c)\, O(r^{-3}), \quad S_\alpha = \mbox{sin}(\Omega 
r/c)\, O(r^{-3}), \end{eqnarray} where $\kappa=e^2a_{0}^{2} 
\Omega^4/(16\pi c^3)$. From (3.28) as well as from (3.25) it is obvious 
that there exist spherical surfaces where either $S_r=0$ or ${\bf S}=0$, 
thus once again confirming the fact that in the stationary states 
described, radiation is absent [20].

Let us describe the vector lines for 
Poynting vector. In spherical system of coordinates we have 
\begin{eqnarray}
\frac{d r}{S_r}\,=\,r\,\frac{d \theta}{S_\theta}\,=\,r\,\mbox{sin}\theta\,
\frac{d \alpha}{S_\alpha}. 
\end{eqnarray}
Last two fractions form integrable combination.  Putting  
$S_\theta$ and $S_\alpha$, from this equality we obtain 
\begin{eqnarray}
\frac{d \theta}{\mbox{sin}\theta\,\mbox{cos}\theta\,}\,=\, 
\mbox{tan}\alpha\,d\alpha,  \end{eqnarray}
which leads to the first integral 
\begin{eqnarray}
|\mbox{tan}\theta\,\mbox{cos}(\alpha-\Omega t)|\,=\,P_1, \quad 
P_1=\mbox{const.} 
\end{eqnarray}
Now we consider first  two fractions. Putting 
$\mbox{cos}(\alpha-\Omega t)\,=\,P_1/\mbox{tan}\theta$ we find
\begin{eqnarray}
\frac{c^2}{\Omega^2}\, \frac{d r}{r^2 \mbox{cos}(\Omega r/c)} =
P_1\, \mbox{sin}\theta\,d \theta. 
\end{eqnarray} 
As we consider the 
region where $r \to \infty$ it is possible  to  factor out the term 
$1/r^2$ from the integrand.  Then we obtain appoximately
\begin{eqnarray}
\frac{c^2}{\Omega^2 r^2}\,\int\, \frac{d r}{ 
\mbox{cos}(\Omega r/c)} = P_1 \int\,\mbox{sin}\theta\,d 
\theta,   \end{eqnarray} 
that leads to 
\begin{eqnarray}
\frac{c^2}{\Omega^2 r^2}\,\mbox{ln} 
|\mbox{tan}(\pi/4 +\Omega r/2c)| \,=\, - P_1\,\mbox{cos}\theta\, + P_2, 
\quad P_2=\mbox{const.}   \end{eqnarray} 
Thus, we built vector lines for Poynting vector in asymptotic region.  

\section{Conclusion}
In the considered soliton model of a hydrogen atom the stability condition 
of spatial stationary motions of electrons in the field of the Coulomb 
center is fulfilled. The existence of this kind of motion had also been
anticipated by Boguslavsky [21] and Chetaev [22]. In particular, due to 
the fulfillment of the nonlinear resonance condition (2.5) the energy 
spectrum of these stationary states coincides with that 
of a hydrogen atom. This fact indicates the role of 
nonlinearity in the formation of extended micro-objects, whose laws of 
evolution agree with quantum mechanics. 

\vskip 5mm
\noindent
{\bf Acknowledgements}\\
B. Saha would like to thank Prof. S. Randjbar-Daemi and ICTP High Energy 
Section for hospitality.

\end{document}